# High Saturation Magnetization, Low Coercivity and Fine YIG Nanoparticles Prepared by Modifying Co-Precipitation Method

S. Hosseinzadeh[a], M. Behboudnia*,[a], L. Jamilpanah[b], M.H. Sheikhi[c], E. Mohajerani[d], K. Tian[e], A. Tiwari[e], P. Elahi[f], S. M. Mohseni*,[b]

[a]Department of Physics, Urmia University of Technology, Urmia, Iran

[b]Faculty of Physics, Shahid Beheshti University, Evin, Tehran, 19839, Iran

[c]Department of Communications and Electronics, School of Electrical and Computer Engineering, Shiraz, Iran

[d]Laser and Plasma Research Institute, Shahid Beheshti University, Evin, Tehran, 19839, Iran

[e]Department of Materials Science and Engineering, University of Utah, Salt Lake City, UT 84112, USA

[f]Department of Mechanical Engineering, University of Utah, Salt Lake City, UT, 84112, USA

*Corresponding author. E-mail Address: m-mohseni@sbu.ac.ir, majidmohseni@gmail.com (Seyed Majid Mohseni). mbehboudnia@gmail.com (Mahdi Behboudnia)

**Abstract**

Nanoparticles with their specific properties newly have drawn a great deal of attention of researchers [1-3]Yttrium iron Garnet magnetic nanoparticles (YIG-NPs) are promising materials with novel applications in microwave, spintronics, magnonics, and magneto-optical devices. However, achieving stable and remarkable magnetic YIG-NPs has been remaining as a great challenge. In this paper, synthesized YIG-NPs by modifying co-precipitation (MCP) method is reported. Structural and magnetic properties of final products are compared to those of the materials prepared by citrate-nitrate (CN) method. Smaller crystals and particle size have been found by MCP method comparing to that of synthesized by CN method. Using a relatively low annealing temperatures for both sets of samples (~700 °C), the final YIG samples prepared by MCP method show more structural purity than those made by CN method. Higher saturation magnetization ($M_s$) and lower coercivity ($H_c$) are observed in MCP YIG sample (23.23 emu/g and 30.1 Oe) than the CN prepared YIG sample (16.43 emu/g and 44.95 Oe). The Curie temperature is measured to be 569 °C for the MCP YIG sample determined from set of $M_s$ measurement at different temperatures ranging from 80-600 K. These findings lead to significant improvement in quality of synthesized (synthetic methods) of YIG-NPs.

**Keywords:** YIG; Co-precipitation; Citrate-nitrate

# 1. Introduction

Nanoparticles with their specific properties newly have drawn great deal of attention of researchers Yittrium iron Garnet (YIG) with chemical composition $Y_3Fe_5O_{12}$ has been abundantly applied in magneto-optical and microwave devices, such as optical insulator, circulators, oscillators and phase shifters [4-6] owing to its narrow linewidth in magnetic resonance, high electrical resistivity, controllable saturation magnetization and large magneto-optical Faraday rotation [7, 8]. Physical responses of YIG specimen are strongly dependent on their microstructure. Moreover, thanks to rapid development in nanotechnology, YIG magnetic nanoparticles (YIG-NPs) have also been investigated variously based on their nanostructure properties [9-13]. Generally, YIG particles can be synthesized using various methods such as co-precipitation, solid-state procedure, auto-combustion and sol–gel methods [14-18]. These methods require high calcination temperature, which decreases monodispersity and also increases particle size. Moreover, the requirement of high heat treatments to prevent reaching any intermediate phases is the distinct disadvantage of these methods. Among above mentioned methods, sol-gel and co-precipitation methods, under some controlled synthesis conditions, have successfully been used for synthesis of nanoparticles. In each method, different parameters like pH, reaction time and temperature and concentration of the materials and solution play a significant role to achieve intended nanoparticles with desired size, shape, and structure [19]. Synthesis of YIG powder via citrate-nitrate (CN) technique, categorized in sol-gel methods, has drawn attentions due to significant advantages such as good mixing texture of precursors, excellent chemical homogeneity of the final products and lower synthesis temperatures compared to other methods [20, 21]. As a comprehensive study, Nguyet et al. [22] synthesized

YIG magnetic NPs by CN at pH=10 followed by annealing at 800 °C for 2h and reported magnetic field and temperature dependencies of magnetization of YIG powder with particle sizes ranging from 45-450 nm.

However, the most prevalent method for synthesizing magnetic NPs is the chemical co-precipitation technique [23-25]. There are several reports on synthesis of YIG magnetic NPs using chemical co-precipitation technique due to its main characteristic features such as simplicity and controllability of the process, low cost, high turnout [26, 27]. It appears likely that the synthesis of fine YIG-NPs at low processing temperatures still has plenty of room to delve in. Generally, NPs of relatively narrow size distributions can be synthesized by co-precipitation technique provided that a short nucleation takes place and be followed by a slower subsequent phase growth. The type of introduced salts, ratio of ions, temperature of the solution, pH, stirring speed and solvents type are the main parameters that must be precisely tuned to yield NPs of desired size with relatively narrow distribution. The purpose of gaining low particle size with high $M_s$ and narrow size distribution has been the topic of many efforts done by authors. Rajendran et al. [28] reported YIG particles with average sizes of 9, 14, 25 nm which were synthesized by co-precipitation followed by modifying the product by chemical treatment. The sample with average particle size of 25 nm showed saturation magnetization ($M_s$) of 20.6 emu/g and no room temperature magnetic moment was reported for the particles having sizes of 9 and 14 nm. Godoi et al. [29, 30] carried out the synthesis of YIG powder by co-precipitation. He and his colleague obtained single phase YIG after annealing the precipitations at 1100 °C and the powder particles were of an average size of about 500 nm. Ristic et al. [31] obtained mainly YIG phase but with a small amount of $YFeO_3$ after annealing at 1200 °C and an ammonia solution in to the aqueous solution of Y- and Fe- nitrates up to pH~10.4. In a study by Rashad et al. [32], the single phase YIG powder was obtained only after annealing the precipitations at 1200 °C via a NaOH solution route. Therefore, getting high saturation magnetization with small size particles and also phase purity of YIG at low temperature synthesis are the broken down purposes for researchers.

In present paper, we report on a low-temperature (700 °C) synthesis of small YIG magnetic NPs via a modifying co-precipitation (MCP) method using DMF as a primary solvent before and during precipitation. We then dissolved precipitation in the solution of citric acid which after washing the precipitates, act as fuel during annealing process. These two modifying approaches lead to formation of small single phase YIG magnetic NPs. We obtained small NPs (~17 nm) with narrower size distribution, higher saturation magnetization (23.23 emu/g) and lower coercivity (30.1 Oe) in comparison with others works. We have also synthesized YIG-NPs by the CN method in pH=1 at the same temperature used for our MCP method and comparing these two methods for final microstructure, size and magnetic properties. Our findings open pathways towards fine production of YIG magnetic NPs with controlled characteristics for distinctive applications.

## 2. Experimental

### 2.1 Materials

Ferric nitrate ($Fe(NO_3)_3 \cdot 9H_2O$), yttrium nitrate ($Y(NO_3)_3 \cdot 6H_2O$), citric acid, ethylene glycol, dimethylformamide (DMF) with 9.98% purity were purchased from Merck.

### 2.2 Preparation of the Samples

Two sets of YIG-NPs were prepared by CN and MCP methods. In CN synthesis, the solution was prepared by dissolving the Y and Fe nitrates in a stoichiometric ratio of Y: Fe = 3:5 in a de-ionized water and a solution of citric acid was added to pH=1. The solution was heated at 80 °C and a gel was obtained after 2 hrs. This gel was dried at 110 °C and then heat treated for 36 h in ambient air at the temperature of 700 °C with a heating rate of 10 °C/min. The color of synthesized powder was brownish-green before heat treatment and turns chartreuse (green-yellow) after heat treatment. For synthesis of YIG-NPs by MCP, yttrium nitrate $Y(NO_3)_3 \cdot 6H_2O$ and iron nitrate $Fe(NO_3)_3 \cdot 9H_2O$ in 3:5 molar ratios were dissolved in DMF to form metal-organic solution. 4-5 drops of ethylene glycol were added as a complexing agent or as a polymerization agent [33, 34]. The solution was continuously stirred using a magnetic stirring bar and stirring speed of 4000 rpm. The mixed hydroxide $3Y(OH)_3 + 5Fe(OH)_3$ was co-precipitated from aqueous solutions up to pH ~ 10.5 by ammonium hydroxide and 25%-ammonia aqueous solution was used as precipitant. The precipitate was stirred for 30 min, centrifuged and then washed with deionized water ethanol. The precipitate was mixed with 2.5 g citric acid and 5.5 mL DI water to reach the pH=2 and continuously stirred at 60 °C to obtain solid precursor and finally heated to 700 °C for 2 hrs. The color of synthesized powder was olive.

## 2.3 Characterization of the samples

Thermal behavior of YIG precursors was determined by thermogravimetric analysis (TG) (model mettle Toledo C1600 analyzer) from ambient temperature to 850 °C in an air atmosphere with a heating rate of 10 °C/min. The crystalline structure of samples was characterized using X-ray diffractometer (STOE-STADI) with Cu Kα (λ = 0.154 nm) radiation. For transmission electron microscopy (TEM) using (LEO 906, Zeiss, 100KV, Germany) and a high-resolution JEOL 2800 S/TEM system was used for performing transmission electron microscopy (HR-TEM). The sample was prepared as follows: a few prepared YIG-NPs were dispersed in absolute ethanol ultra-sonication for 10 min and was dropped over carbon-coated copper grid and dried at room temperature. Room temperature magnetization measurement were done with vibrating sample magnetometer (VSM, Meghnatis Daghigh Kavir Co.) and temperature dependent magnetization measurements were done using a Microsense FCM-10 vibrating sample magnetometer (VSM), The volume-average diameter and size distribution of YIG magnetic NPs was measured by DLS (Shimadzu UV-1800 spectrophotometer). For DLS measurement, the YIG-NPs dispersed in absolute ethanol as a suitable solvent with appropriate dispersing agents and sonicate it for 10 min.

## 3. Results and discussion

### 3.1. X-Ray Diffraction (XRD) Study

Figure 1 shows XRD pattern of YIG for CN and MCP samples annealed at 700 °C at the room temperature. Prepared samples from CN contain YIG phase, along with peaks that can be attributed to maghemite (ɣ-$Fe_2O_3$), hematite (α-$Fe_2O_3$) and orthoferrite ($YFeO_3$). These residual phases may be formed due to insufficient sintering time or temperature. The proposed crystallization process can be described by following reactions [35, 36]:

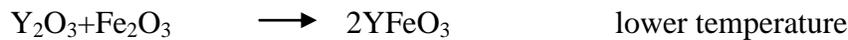
$Y_2O_3 + Fe_2O_3 \longrightarrow 2YFeO_3$      lower temperature

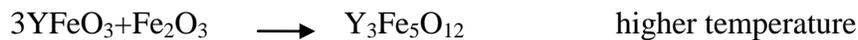
$3YFeO_3 + Fe_2O_3 \longrightarrow Y_3Fe_5O_{12}$      higher temperature

Prepared sample by MCP method (annealed at 700 °C) completely contains YIG phase with very negligible contribution of intermediate phases. Therefore, crystallization occurs at lower temperature which can be happened because of using DMF as the solvent instead of water of precursors before precipitation process takes place. DMF as a polar solvent helps the diffusion and increases effective contact of the reactant molecules compared with water, which could well disperse the ions and surround each ion during the precipitation process [37]. DMF as an organic solvent and citric acid as a complexing agent, helps to bring the yttrium and iron cations closer to each other. If the cations are close enough to each other, crystallization of oxide phase requires a much lower amount of energy in the system which will assist the formation of crystalline structure in lower temperature. Using Miller indices, the unit cell parameter was determined to be $a_0$= 12.404 Å which is larger than the bulk value of $a_0$ = 12.3774 Å (JCPDS 33-693). Similarly, an increase in $a_0$ is observed for, $a_0$= 12.416 Å. Crystal size was calculated by fitting of all the peaks and taking average of all the sizes obtained by Scherrer's equation. The average crystal size obtained from CN and MCP methods are 23 nm and 17 nm, respectively.

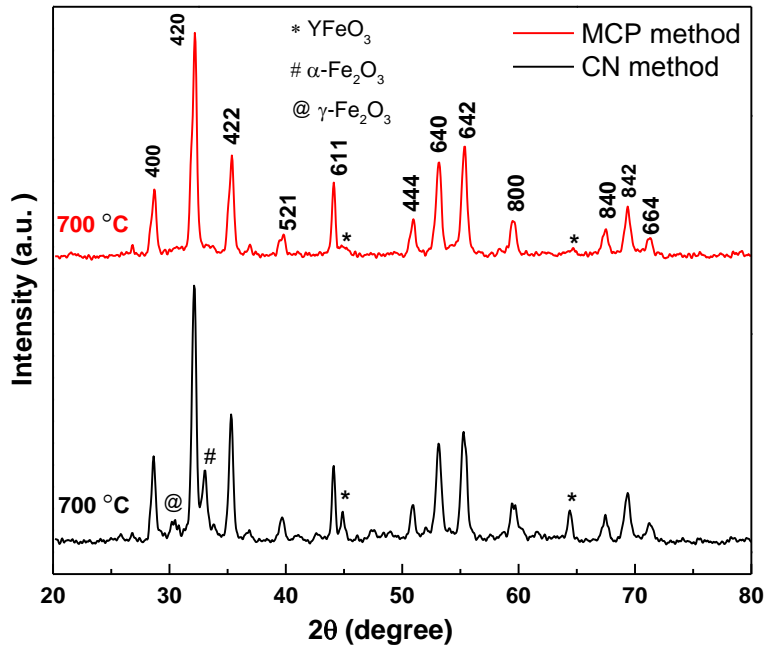

Figure 1: XRD pattern of (a) CN YIG-NPs and (b) MCP YIG-NPs at room temperature

## 3.2. Thermogravimetric Analysis (TGA)

As observed in XRD data, the sample prepared by MCP method is more crystalline and simultaneously has a lower crystal dimensions in comparison with the one prepared by CN method. In order to investigate the decomposition performance under air atmosphere, TGA technique is carried out for MCP prepared YIG sample. Figure 2 shows TGA curve of this sample and the steps of decomposition process in MCP method are shown. The TGA curve showed an overall weight loss of 58.6%. The weight loss of <10 % in the temperature range of 100-190 °C is due to evaporation of residual water molecules from the gels. The second weight loss which is about 63% of the overall weight loss occurs in the temperature range of 190-500 °C and can be considered as decomposition of remaining organics or oxidation of residual carbons. The weight loss in the temperature range of 500-690 °C is associated with crystallization of $YFeO_3$ and the last weigh loss occurs at the temperature range of 690-760 °C which corresponds to formation and crystallization of YIG. It is granted that the starting temperature of decomposition is 190 °C and YIG formation completes between 690 to 760 °C. We noticed that the crystallization temperature is considerably decreased when organic acids are used. These molecules contain carbon which can act as fuel in the annealing process [16]. This means that higher local temperature assists the crystallization of compound. We used minimum amount of citric acid at 700 °C in MCP method to get smallest particle size. On the other hand, because of low annealing temperature we chose pH of 1 for the prepared sample by CN method among pH of 1, 2 and 3 reported in literatures in order to obtain complete phases of YIG since it is reported that by reducing pH, annealing temperature will decrease as well. The details of CN method and related TG curves can be found in referenced literature [14, 38].

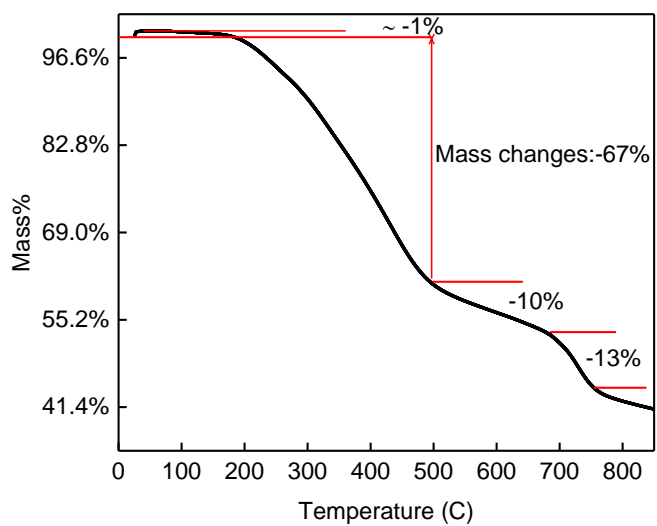

Figure 2: TGA curve of YIG particles prepared by MCP method.

### 3.3. Morphology and size of particles

Figure 3 shows TEM images of YIG-NPs prepared by CN and MCP methods. High resolution TEM (HRTEM) and SAED pattern for MCP YIG sample is presented in the figure 3a corresponds to YIG-NPs synthesized by CN method. It can be seen that particles are aggregated and exhibit irregular shapes without shaped borders. Figure 3b consequently determines that the size and line shape distribution of the particles. Because of the observed aggregation in TEM images of this sample the DLS results show a much larger size for it. Agglomeration of NPs in the absence of surfactants is common[39]. When we disperse magnetic nanoparticles in solution (ethanol) for DLS measurement, the magnetic interaction among the NPs larger than 20 nm in diameter is large enough to dominate the Brownian force among them while the smaller NPs remain stable in the solvent[1, 40]. Thus, considering the tendency of agglomeration of YIG NPs prepared by CN (their size are bigger than 20nm) in spherical aggregates the size was obtained by DLS is different by the size obtained by TEM and XRD. Figure 3c illustrates TEM photograph of YIG-NPs synthesized by MCP method showing moderate clustering of particles and we can observe some small aggregations, which are composed of primary particles with the size distribution of 10-20 nm, matches the result of DLS measurement. Figure 3d describes in accordance with the calculated values for crystal size by the Scherrer's equation, the mean diameter of YIG-NPs prepared by MCP is 17 nm, which agrees with the values calculated from XRD. The YIG magnetic NPs prepared by MCP are nearly single crystal and the TEM analysis is quite consistent with the size distribution analyses mentioned above. The inset of Figure 3d shows the SAED pattern of YIG magnetic NPs synthesized by MCP, and annealed at 700 °C which contains single crystals. The bright spots are indexes to the (642), (422), (444), (644), (420), and (842) of the YIG-NPs. Figure 3e shows the lattice fringe image. The fringe spacing is measured to be 0.31 nm, 0.22 nm and 0.25 nm which correspond to the (400), (521) and (422) crystallographic plane of YIG and we show only the 0.22 nm space in the figure.

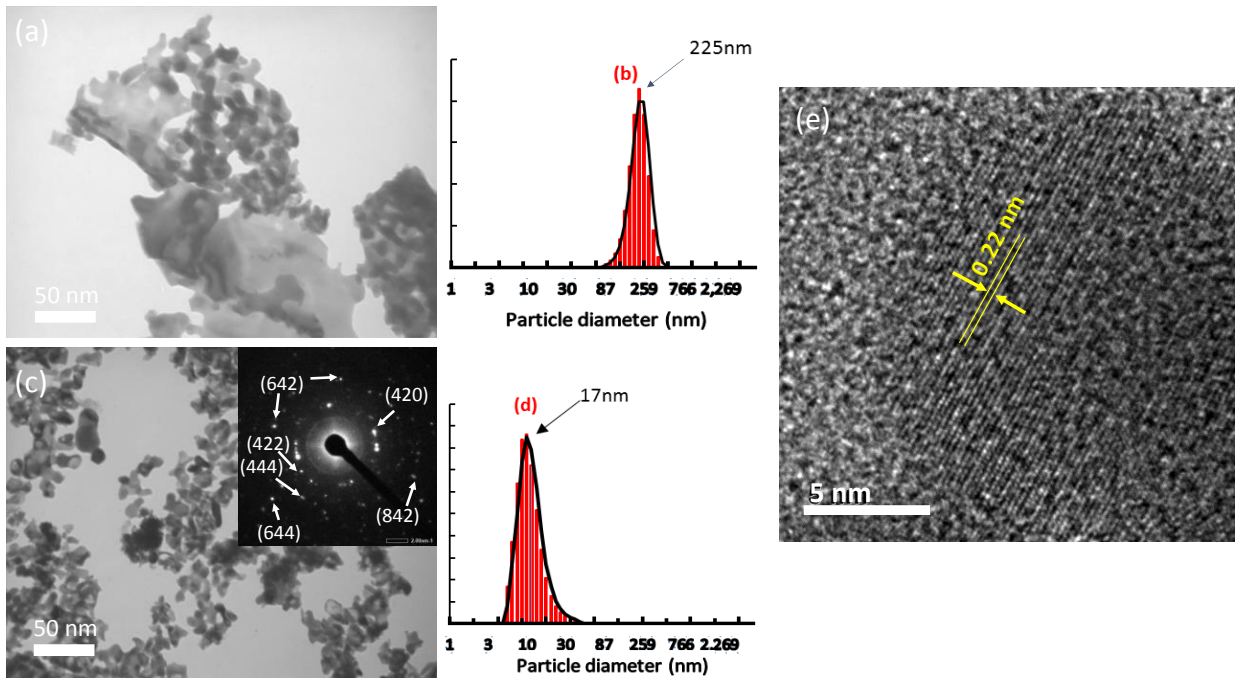

Figure 3: a) TEM image, b) size distribution from DLS of YIG-NPs synthesized by CN, c) TEM image, d) size distribution from DLS of YIG-NPs synthesized by MCP method, e) HRTEM of YIG-NPs synthesized by MCP method.

## 3.4. Magnetic Hysteresis Loops

In order to see the effect of synthesis method on the magnetic properties of YIG magnetic NPs, the magnetic response of the samples in the magnetic field was analyzed and evaluated by VSM at different temperatures. Figure 4 shows the hysteresis loops recorded at room temperature for both samples. The $M_s$ of CN and MCP prepared samples are 16.43 and 23.23 emu/g, respectively. It is shown that YIG particles prepared by CN method have a lower value of Ms than those prepared by MCP method, which is more than the $M_s$ reported in literatures [26, 28]. The lower measured value of magnetic saturation in CN method originated from existence of intermediate phases such as $α/ɣ\text{-}Fe_2O_3$ which is a weak ferromagnetic compound and $YFeO_3$ which is antiferromagnetic compound [41, 42]. The presence of mentioned intermediate phases is owing to the insufficient annealing temperature which exacerbates YIG transition in CN method. Moreover, magnetic NPs synthesized by CN method have larger $H_c$ and $M_r$ than MCP prepared NPs (inset of Figure 4). We attribute this phenomenon to the existence of secondary phases and increment of magnetically disordered in structures [22, 43]. The data of $H_c$, $M_s$ and $M_r$ for both sets of samples can be seen in Table 1.

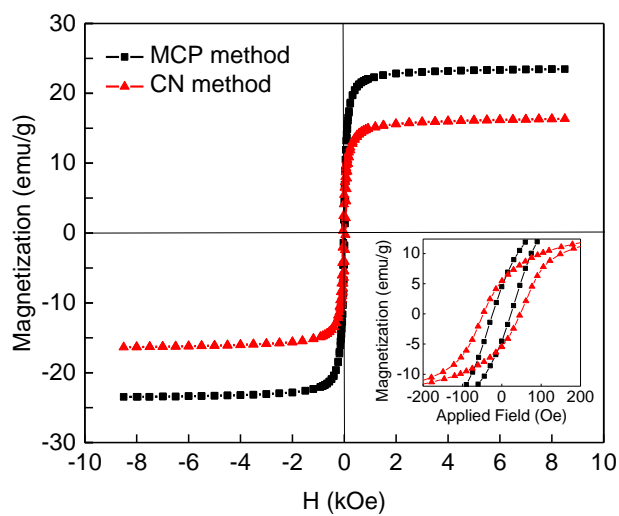

Figure 4: Magnetic hysteresis loops of YIG particles prepared by CN and MCP methods.

Table 1: Particle sizes obtained from XRD, TEM and DLS and $M_s$, $H_c$ and $M_r$ observed from VSM measurements.

| Sample | Crystal size from XRD (nm) | Particle size from TEM (nm) | Particle size from DLS (nm) | $M_s$ (emu/g) | $H_c$ (Oe) | $M_r$ (emu/g) |
|---|---|---|---|---|---|---|
| CN | 23 | > 20 nm | 100-389 | 16.43 | 44.95 | 5.49 |
| MCP | 17 | ~20 | 6-26 | 23.23 | 30.1 | 4.52 |

To further study the magnetic properties of the YIG-NPs prepared via MCP method we measured temperature dependent $M_s$. The magnetization loops M (H) in the temperatures ranging from 80 to 600 K were measured using VSM. The loops at 80, 100, 150, 200, 250 and 300 K are presented in Figure 5a and those ranging from 350 to 600K are presented in Figure 5b. A common feature of the loops at different temperatures is the magnetization approaches to saturation above 2000 Oe. At 80 the $M_s$ is 4.3 $\mu_B$/f.u., being 86% of theoretical value of YIG bulk which is 5 $\mu_B$/f.u [44]. Based on the relation $M_s^{ex\,primental} = M_s^{bulk}[(D/2-t)/D/2]^3$ for core-shell morphology where D is the particle diameter and t the surface layer thickness, a value of t= 0.4 nm is obtained according to mean particle size of 17 nm. At the inset of Figure 5a, it can be seen that the coercivity decreases as the temperature decreases. Also in the inset of the Figure 6 there is a phase change at temperatures higher than 500 K. At higher temperatures the coercivity vanishes which indicates a super paramagnetic behavior of the sample.

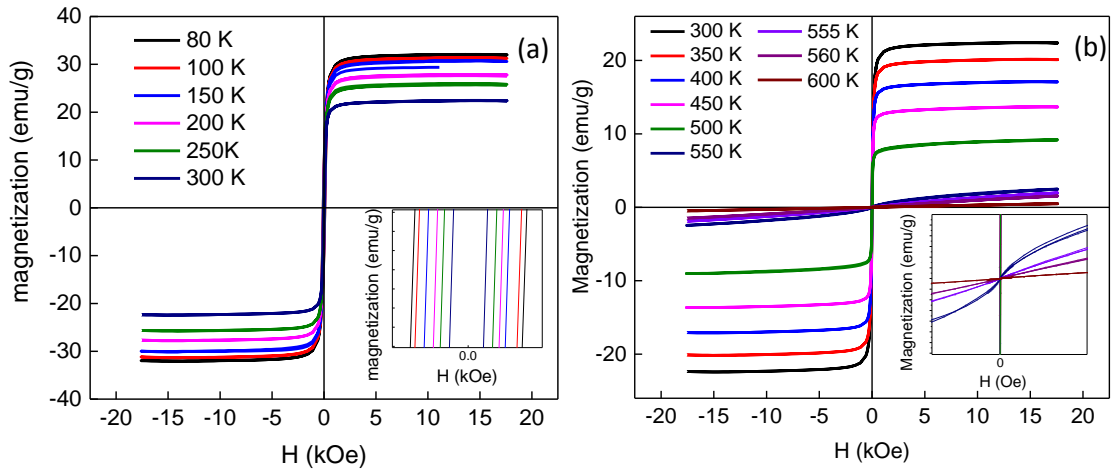

Figure 5: a) Hysteresis loops measured at 80, 100, 150, 200, 250 and 300 $K$ for the MCP YIG-NPs. The inset shows the magnified region around the origin and b) Magnetization loops measured at temperatures between 350 and 560 $K$ for the MCP YIG-NPs. The inset shows the magnified region around the origin.

The temperature dependence of the saturation magnetization for MCP YIG sample is exhibited in Figure 6. The experimental data was fitted using Bloch's equation:

$$M_T = M(0)[1-(\frac{T}{T_0})^b]$$

Where, $M_0$ is the saturation magnetization at 0 K, $T_0$ is the temperature at which the $M_s$ reduces to zero (Curie temperature, $T_C$) and the exponent b is the known Bloch's exponent. The fitted magnetic parameters are shown in the Figure 6. From this fit the Curie temperature of MCP YIG-NPs is found to be 569 K, not far from previous achievements (560 K) [1,2]. The Bloch exponent obtained to be 2.45 from the fitting process. In our samples, it is observed from the fitting parameters that the modified exponent (b) is larger than the exponent for bulk materials (3/2). For NPs, it is expected that as the size of the particle increases, b approaches to that of the bulk materials. The deviation from the original Bloch's law could be attributed to the surface effects and increment of the disordered states [6, 7]. In the case of our sample the particle size is small (17 nm) and this effect is expected to occur.

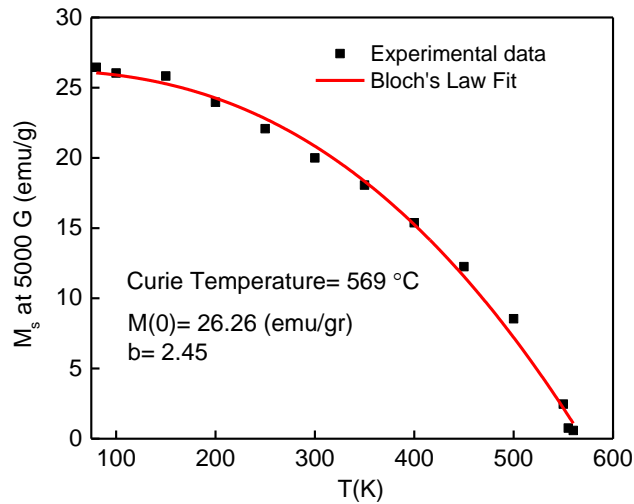

Figure 6: Saturation magnetization, $M_s$ versus temperature, T for MCP sample. Solid line shows fitting with Bloch's equation.

Coercivity of the MCP sample at different temperatures is plotted in Figure 8. At approximately 300 K sample losses its coercivity. It is known from theoretical model that, for a given particle size, $H_c$ decreases with increasing temperature following the Kneller's formula [22, 45]. In Kneller's formula $H_c$ decreases by increasing temperature proportional to $\sqrt{T}$ while it can be seen in Figure 8 that the reduction of $H_c$ is nearly linear with T. Different particle sizes and different anisotropy alignments of particles can cause deviations from Kneller's formula but yet we see the procedure of reductive behavior in the figure.

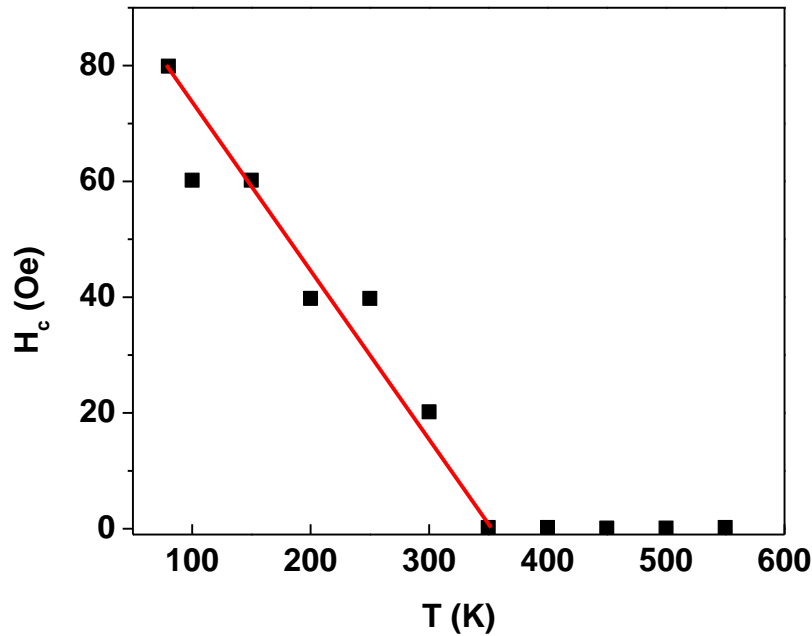

Figure 7: Coercivity, $H_c$ versus temperature, T for MCP sample. Red line is a guide to eye and dots are experimental data

**Conclusion**

MCP method yielded to finer YIG-NPs with higher saturation magnetization and lower coercivity. Minimum amount of citric acid acts as a complexing agent and also as fuel in annealing process in MCP method. Furthermore, citric acid takes a prominent role in decreasing crystallization temperature beside using DMF as polar solvent which both in their turns result in finer particles with relatively high saturation magnetization compared with the chemical CN method. YIG MCP particles show saturation magnetization of 23.23 emu/gr and the Curie temperature of 569 °C. The variation of saturation magnetization with temperature follows a $T^{2.4}$ power-law which is attributed to the small size of prepared YIG-NPs. Our findings can be used in synthesizing and achieving desired size and characteristics of YIG magnetic NPs.

**Acknowledgments**

S.M. Mohseni acknowledges support from Iran Science Elites Federation (ISEF), Iran Nanotechnology Initiative Council (INIC) and Iran's National Elites Foundation (INEF). A.T. wants to thanks US NSF for support through grant No. 1407650.